\documentclass[aps, prd, amsmath, floats, floatfix, twocolumn, nofootinbib, superscriptaddress]{revtex4-1}
\usepackage[pdftex]{graphicx}
\usepackage{color}
\usepackage{latexsym}
\usepackage[colorlinks]{hyperref}
\usepackage{amsthm,amsmath,amssymb}
\usepackage{mathrsfs}
\usepackage{makecell}
\usepackage{float}
\usepackage{lipsum}

\newcommand{\be}{\begin{eqnarray}}
\newcommand{\ee}{\end{eqnarray}}

\newcommand{\apjl}{Astrophys.\ J.\ }
\newcommand{\apjs}{Astrophys.\ J. Suppl.\ }
\newcommand{\aap}{Astron.\ Astrophys.\ }

\newcommand{\mnras}{Mon.\ Not.\ R.\ Astron.\  Soc.\ }
\newcommand{\pasj}{Publ.\ Astron.\ Soc.\ Japan\ }

\newcommand{\ssr}{Space Sci.\ Rev.\ }

\newcommand{\tabincell}[2]{\begin{tabular}{@{}#1@{}}#2\end{tabular}}
\begin{document}

\title{Impact of the returning radiation on X-ray reflection spectroscopy measurements:\\the case of Galactic black holes}

\author{Kexin~Huang}
\affiliation{Center for Astronomy and Astrophysics, Center for Field Theory and Particle Physics, and Department of Physics,\\
Fudan University, Shanghai 200438, China}

\author{Honghui~Liu}
\email[E-mail: ]{honghui.liu@uni-tuebingen.de}
\affiliation{Institut f\"ur Astronomie und Astrophysik, Eberhard-Karls Universit\"at T\"ubingen, D-72076 T\"ubingen, Germany}

\author{Cosimo~Bambi}
\email[E-mail: ]{bambi@fudan.edu.cn}
\affiliation{Center for Astronomy and Astrophysics, Center for Field Theory and Particle Physics, and Department of Physics,\\
Fudan University, Shanghai 200438, China}
\affiliation{School of Natural Sciences and Humanities, New Uzbekistan University, Tashkent 100007, Uzbekistan}

\author{Javier~A.~Garc{\'\i}a}
\affiliation{NASA Goddard Space Flight Center, Greenbelt, MD 20771, United States}
\affiliation{Cahill Center for Astronomy and Astrophysics, California Institute of Technology, Pasadena, CA 91125, United States}

\author{Zuobin~Zhang}
\affiliation{Center for Astronomy and Astrophysics, Center for Field Theory and Particle Physics, and Department of Physics,\\
Fudan University, Shanghai 200438, China}

\begin{abstract}
The effect of the returning radiation has long been ignored in the analysis of the reflection spectra of Galactic black holes and active galactic nuclei and only recently has been implemented in the {\tt relxill} package. Here we present a study on the impact of the returning radiation on the estimate of the parameters of Galactic black holes. We consider high-quality \textit{NuSTAR} spectra of three Galactic black holes (GX~339--4, Swift~J1658.2--4242, and MAXI~J1535--571) and we fit the data with the lamppost model in the latest version of {\tt relxill}, first without including the returning radiation and then including the returning radiation. We do not find any significant difference in the estimate of the parameters of these systems between the two cases, even if all three sources are fast-rotating black holes and for two sources the estimate of the height of the corona is very low, two ingredients that should maximize the effect of the returning radiation. We discuss our results and the approximations in {\tt relxill}.   
\end{abstract}

\maketitle


\section{Introduction}

Blurred reflection features are commonly detected in the X-ray spectra of Galactic black holes and active galactic nuclei~\cite{2007MNRAS.382..194N} and are generated by the illumination of a cold disk by a hot corona~\cite{1989MNRAS.238..729F,2013Natur.494..449R}. In the presence of high-quality data and with the correct astrophysical model, the analysis of these blurred reflection features is a powerful tool for studying the accretion process in the strong gravity region of black holes, measuring black hole spins (and it is currently the only mature technique for measuring the spins of supermassive black holes), and testing Einstein's theory of General Relativity in the strong field regime~\cite{2021SSRv..217...65B,2023ApJ...946...19D,2024ApJ...969...40D,2018PhRvL.120e1101C,2019ApJ...875...56T,2021ApJ...913...79T}.

The state-of-the-art in reflection modeling is currently represented by the non-relativistic reflection models {\tt xillver}~\cite{2013ApJ...768..146G} and {\tt reflionx}~\cite{2005MNRAS.358..211R} and by the relativistic models {\tt relxill}~\cite{2013MNRAS.430.1694D,2014ApJ...782...76G}, {\tt reltrans}~\cite{2019MNRAS.488..324I}, {\tt reflkerr}~\cite{2008MNRAS.386..759N,2019MNRAS.485.2942N}, and {\tt kyn}~\cite{2004ApJS..153..205D}. There is even the model {\tt relxill\_nk}~\cite{2017ApJ...842...76B,2019ApJ...878...91A}, which is an extension of the {\tt relxill} package to non-Kerr spacetimes. Despite the efforts over the past 20~years to develop more and more advanced reflection models, these models still rely on a number of simplifications, so caution is necessary when we analyze high-quality data because we may get very precise but not very accurate measurements.

Simplifications in the calculations of synthetic reflection spectra can be roughly grouped into four classes: $i)$ simplifications in the description of the accretion flow, $ii)$ simplifications in the description of the corona, $iii)$ simplifications in the calculations of the non-relativistic reflection spectra, and $iv)$ relativistic effects that are neglected or not properly taken into account. There are already a number of publications investigating the impact of these simplifications on current measurements of accreting black holes, as well as efforts to remove some of these simplifications and develop more advanced reflection models; see, for instance, Refs.~\cite{2011MNRAS.414.1269W,2012MNRAS.424.1284W,2014PhRvD..89l7302B,2017MNRAS.472.1932G,2020PhRvD.101d3010Z,2020PhRvD.101l3014C,2020ApJ...899...80A,2021PhRvD.103j3023A,2021ApJ...913..129T,2021ApJ...923..175A,2022ApJ...938...53S,2022MNRAS.514.3246J,2022MNRAS.517.5721M,2024PhRvD.110d3021L,2024arXiv241200349L} for an incomplete list of these studies.

The {\it returning radiation} is the radiation that is emitted by the accretion disk and returns to the accretion disk because of the strong light bending near a black hole. The effect of the returning radiation on the reflection spectrum of an accretion disk was first studied in Ref.~\cite{1997MNRAS.288L..11D} and further investigated in Refs.~\cite{2016ApJ...821L...1N,2018MNRAS.477.4269N,2021ApJ...910...49R,2022MNRAS.514.3965D,2023EPJC...83..838R} with different approximations. All these studies agree that the effect is important only for fast-rotating black holes (with a spin parameter $a_* \gtrsim 0.9$, assuming that the inner edge of the disk is located at the innermost stable circular orbit) and when the corona is compact and close to the black hole (in the case of a lamppost corona\footnote{In the lamppost geometry, the corona is a point-like, static, and isotropic source along the black hole spin axis. This is a very simple coronal model (there is only one parameter, the height of the coronal $h$), but it can be used to approximate the basis of a jet, which could indeed act as a hot corona~\cite{Markoff:2005ht,Kara:2019zad}.}, $h \lesssim 5$~$r_{\rm g}$, where $h$ is the height of the corona and $r_{\rm g} = G_{\rm N}M/c^2$ is the gravitational radius of the compact object). Recently, the effect of the returning radiation has been implemented in the {\tt relxill} package in the calculation of the emissivity profile and without changing the non-relativistic reflection spectrum~\cite{2022MNRAS.514.3965D}. As shown in Ref.~\cite{2024ApJ...965...66M}, such an approximation works better for low values of the black hole spin parameter, high values of the height of the corona, and extremely high value of the ionization parameter. It is worse for fast-rotating black holes and lower values of the coronal height and ionization parameter. Indeed the reflection spectrum is determined by the incident spectrum illuminating the disk: the X-ray spectrum of the direct radiation from the corona is close to a power law with a high-energy cutoff while the spectrum of the returning radiation is a reflection spectrum. In general, the picture is even more complicated because even the returning radiation of the thermal radiation from the disk can contribute to generate the reflection spectrum~\cite{2024ApJ...976..229M}. A recent study of IXPE data of Cygnus~X-1 in the soft state has shown that the polarization signal in the soft X-ray band is dominated by the returning reflection emission~\cite{2024ApJ...969L..30S}.

In the present paper, we want to figure out the impact of the returning radiation on the measurement of the parameters of a reflection model. We use the latest version of {\tt relxill}~\cite{2022MNRAS.514.3965D}, where the flag {\tt switch\_returnrad} controls the returning radiation: if ${\tt switch\_returnrad} = 0$ the spectrum is calculated without including the returning radiation and if ${\tt switch\_returnrad} = 1$ the spectrum is calculated including the effect of the returning radiation on the emissivity profile. For our study, we select high-quality \textit{NuSTAR} spectra of three Galactic black holes (GX~339--4, Swift~J1658.2--4242, and MAXI~J1535--571). From previous analyses of these spectra, we know that there are strong blurred reflection features, the data can be fit well with the lamppost corona model, and all three sources are fast-rotating black holes. We repeat the analysis of these spectra, first with ${\tt switch\_returnrad} = 0$ and then with ${\tt switch\_returnrad} = 1$. In the case of GX~339--4, previous studies found a coronal height $h \sim 10$~$r_{\rm g}$, but the quality of the data is very good, so we can still expect to see some differences in the estimate of the model parameters when we turn the returning radiation off and on. For Swift~J1658.2--4242 and MAXI~J1535--571, previous analyses suggest that the height of the corona is very low, $h < 3$~$r_{\rm g}$. Despite the high-quality of these spectra and the specific properties of these sources, which should maximize the effects of the returning radiation on the observed relativistically blurred reflection spectra, we do not find any significant difference in the estimate of the model parameters when we include and we do not include the returning radiation.

The manuscript is organized as follows. 
In Section~\ref{observation}, we introduce the sources and the observations analyzed in this paper.
In Section~\ref{s-red} and Section~\ref{model}, we describe the data reduction and analysis, respectively. Discussion and conclusions are reported in Section~\ref{discussion}.

\begin{figure}[t]
	\centering
	\includegraphics[width=0.92\linewidth]{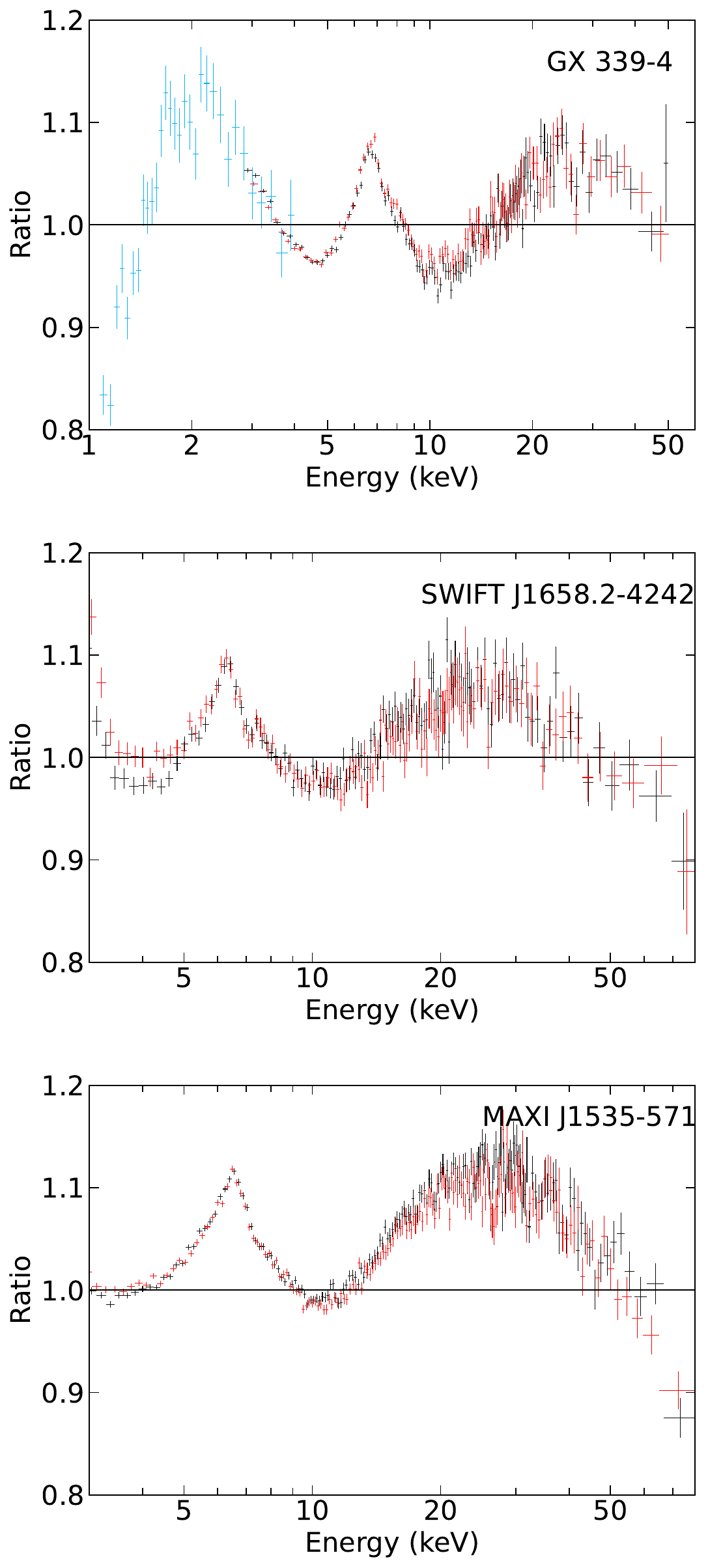}
    \vspace{0cm}
	\caption{Data to best-fit model ratio plots for an absorbed power-law model for our three sources (absorbed disk blackbody and power-law model for the spectrum of GX 339--4). The blue crosses are for \textit{Swift}/XRT data and the black and red crosses are for \textit{NuSTAR}/FPMA and \textit{NuSTAR}/FPMB data, respectively.
	\label{ironline}}
\end{figure}

\section{Sources and Observations}
\label{observation}

In this study, we investigate the spectral properties of three Galactic black holes, chosen for their distinct characteristics and well-documented observations. A summary of the sources and corresponding observational details is presented in Table \ref{observation_table}.

\begin{table*}[t]
	\centering
	\renewcommand\arraystretch{1.5}
	\caption{List of the observations analyzed in this paper.}
	\label{observation_table}
	\begin{tabular}{llcccc}
		\hline
		 & Source & \hspace{0.2cm} Mission \hspace{0.2cm} & \hspace{0.2cm} Observation ID \hspace{0.2cm} & \hspace{0.2cm} Observation Date \hspace{0.2cm} & \hspace{0.2cm} Exposure (ks) \hspace{0.2cm} \\
		\hline
		 & GX~339--4 & \textit{NuSTAR}  & 80001015003    & 2015 March 11 & 30.0      \\
		 &               & \textit{Swift}   & 00081429002     & 2015 March 11 & 1.9  \\
		\hline
         & Swift~J1658.2--4242 & \textit{NuSTAR} & 90401307002 & 2018 February 16 & 33.3\\
            \hline
         & MAXI~J1535--571 & \textit{NuSTAR} & 90301013002 & 2017 September 7 & 10.3 \\
            \hline
	\end{tabular}\\
\end{table*}

\subsection{GX~339--4}

GX~339--4 is a low-mass X-ray binary known for its frequent outbursts. It was discovered in 1973 by \citet{1973ApJ...184L..67M}. On March 11, 2015, the source was simultaneously observed with \textit{NuSTAR} and \textit{Swift}. The data from this observation were first analyzed by \citet{2016ApJ...821L...6P}. The authors found the black hole spin parameter $a_* \sim 0.95$ and the height of the corona $h \sim 9$~$r_{\rm g}$.

\subsection{Swift~J1658.2--4242}

Swift~J1658.2--4242 was first detected by \textit{Swift}/BAT on February 16, 2018. The X-ray spectral and timing properties indicated that the compact object was a black hole. The source was also observed by \textit{NuSTAR} on the same day the outburst was first detected, with an exposure time of 33.3~ks. The analysis of this observation was reported in \citet{2018ApJ...865...18X} and the spectral analysis suggested a high value of the black hole spin parameter and a low value of the height of the corona. In our study, only the \textit{NuSTAR} data are considered due to the very low quality of the \textit{Swift} spectrum.

\subsection{MAXI~J1535--571}

MAXI~J1535--571 was discovered as a hard X-ray transient by \textit{MAXI}/GSC \citep{2017ATel10708....1N} and \textit{Swift}/BAT \citep{2017ATel10731....1K} on September 2, 2017. Subsequent X-ray and radio monitoring confirmed its classification as a black hole X-ray binary \citep{2017ATel10711....1R}. On September 7, 2017, the source was observed by \textit{NuSTAR} and the data were analyzed by \citet{2018ApJ...852L..34X}. The authors found a black hole spin parameter $a_* > 0.84$ and a corona height $h \sim 7$~$r_{\rm g}$.

\section{Data reduction}\label{s-red}

The spectra used in our study were previously analyzed using the lamppost model in \citet{2016ApJ...821L...6P}, \citet{2018ApJ...865...18X}, and \citet{2018ApJ...852L..34X}, but with an earlier version of \texttt{relxill} that did not include the effect of returning radiation. We adopt the data reduction procedures outlined in those works with the modification that we apply the optimal binning algorithm described in \citet{2016A&A...587A.151K} using the \texttt{ftgrouppha} task. We summarize the main steps of the reduction process below.

\subsection{\textit{NuSTAR}}

For the \textit{NuSTAR} observations, we used the tool \texttt{nupipeline} v0.4.9 to clean the event files for both the FPMA and the FPMB detectors, employing the calibration database CALDB v20230307. The source spectra were extracted from circular regions centered on the sources, with a radius of 150~arcsec for GX~339--4 and Swift~J1658.2--4242, and 180~arcsec for MAXI~J1535--571. The response files were generated using the \texttt{nuproducts} task from NuSTARDAS v2.1.2. Background spectra were extracted from regions far from the sources to minimize the contribution from the sources.

\subsection{\textit{Swift}}

For the XRT/\textit{Swift} observation, the raw data were processed using the \texttt{xrtpipeline} v0.13.7 script to generate cleaned event files. We extracted the source spectra from an annular region with an inner radius of 25~arcsec and an outer radius of 45~arcsec to avoid the possible pile-up region during spectra extraction. The background spectra were extracted from a source-free region. Ancillary response files (ARFs) were generated using the \texttt{xrtmkarf} task.

\section{Data Analysis}\label{model}

We use XSPEC v12.12.1 \citep{1996ASPC..101...17A} to analyze the X-ray spectra. The data are grouped using the optimal binning algorithm described by \citet{2016A&A...587A.151K}.

\subsection{{Reflection features}}

For each source, we begin by fitting the spectrum with an absorbed power-law model. In the case of GX~339--4, we also include a disk blackbody spectrum. In XSPEC language, the model for GX~339--4 reads

\begin{equation} {\tt tbabs} \times ({\tt diskbb} + {\tt powerlaw}). \end{equation}

\noindent {\tt tbabs} describes the Galactic absorption \cite{2000ApJ...542..914W} and has one free parameter: the hydrogen column density, $N_{\rm H}$. {\tt diskbb} describes the thermal radiation from the accretion disk~\cite{1984PASJ...36..741M} and has two free parameters: the inner disk temperature, $T_{\rm in}$, and the normalization of the component, Norm. {\tt powerlaw} accounts for the coronal emission and has two free parameter: the photon index, $\Gamma$, and the normalization of the component, Norm. For the other two sources, we use the absorbed power-law model

\begin{equation} {\tt tbabs} \times {\tt cutoffpl}, \end{equation}

\noindent where {\tt cutoffpl} is a power-law component with an exponential high-energy cutoff, so the free parameters are the photon index, $\Gamma$, the high-energy cutoff, $E_{\rm cut}$, and the normalization of the component, Norm.

\begin{table*}[t]
	\centering
	\renewcommand\arraystretch{1.5}
	\caption{List of the models used for every sources in our study.}\label{t-model}
	\label{models}
	\begin{tabular}{lc}
	\hline
	Source & Model \\
	\hline
	GX~339--4 & {\tt tbabs} $\times$ ({\tt kerrbb} + {\tt relxilllpCp})     \\
	\hline
        Swift~J1658.2--4242 \hspace{0.3cm} & {\tt tbabs} $\times$ ({\tt relxilllpCp} + {\tt gaussian})      \\
        \hline
        MAXI~J1535--571 & {\tt tbabs} $\times$ ({\tt diskbb} + {\tt relxilllpCp} + {\tt xillverCp}) \\
        \hline
	\end{tabular}\\
\end{table*}

The plots of the data to best-fit model ratios of the three sources are shown in Fig.~\ref{ironline}. In all cases, we observe prominent relativistic reflection features, including a broadened iron line in the 5-8~keV range and a Compton hump peaking around 20~keV. For Swift~J1658.2--4242, we also see absorption features around 7~keV, but the broadened iron line is still clearly visible.

\subsection{Reflection analysis}

To model the relativistic reflection features, we employ the lamppost model {\tt relxilllpCp} from the {\tt relxill} v2.0 family~\cite{2022MNRAS.514.3965D}. Initially, we exclude the returning radiation by setting the {\tt switch\_returnrad} parameter to 0. We note that {\tt relxilllpCp} with free reflection fraction $R_{\rm f}$ describes both the reflection spectrum from the disk and the continuum from the corona, so we do not need {\tt powerlaw} or {\tt cutoffpl}, which were used in the previous subsection.

For GX~339--4, we replace {\tt diskbb} used in the previous subsection with {\tt kerrbb}~\cite{2005ApJS..157..335L} and we fit the data with the model

\begin{equation} {\tt tbabs} \times ({\tt kerrbb} + {\tt relxilllpCp}). \end{equation}

\noindent For Swift~J1658.2--4242, we include {\tt gaussian} to model a narrow emission line around 6.3~keV, which is likely due to reflection from distant material. We fit the spectrum of Swift~J1658.2--4242 with the model

\begin{equation} {\tt tbabs} \times ({\tt relxilllpCp} + {\tt gaussian}). \end{equation}

\noindent For MAXI~J1535--571, we need to add {\tt diskbb} to model a weak thermal component and {\tt xillverCp} to account for a narrow Fe~K$\alpha$ line from distant material

\begin{equation} {\tt tbabs} \times ({\tt diskbb} + {\tt relxilllpCp} + {\tt xillverCp}). \end{equation}

\noindent Table~\ref{models} summarizes the final XSPEC models used to fit the spectrum of every source. After the first round of fit without including the returning radiation, we refit all spectra by setting {\tt switch\_returnrad} to 1 in order to include the effect of the returning radiation in the emissivity profile of the disk.

\begin{figure*}
	\centering
	\caption{Best-fit model and residuals for GX~339--4 when we turn the returning radiation off (left panel) and on (right panel). The black lines represent the total model, the blue dashed lines are for the thermal component ({\tt kerrbb}), and the red dashed lines are for the reflection spectrum from the disk and the continuum from the corona ({\tt relxilllpCp}). In the lower quadrants, the blue crosses are for \textit{Swift}/XRT data and the black and red crosses are for \textit{NuSTAR}/FPMA and \textit{NuSTAR}/FPMB data, respectively.}
	\label{gx339_model_ra_plot}
	\vspace{0cm}
	\includegraphics[width=0.99\linewidth]{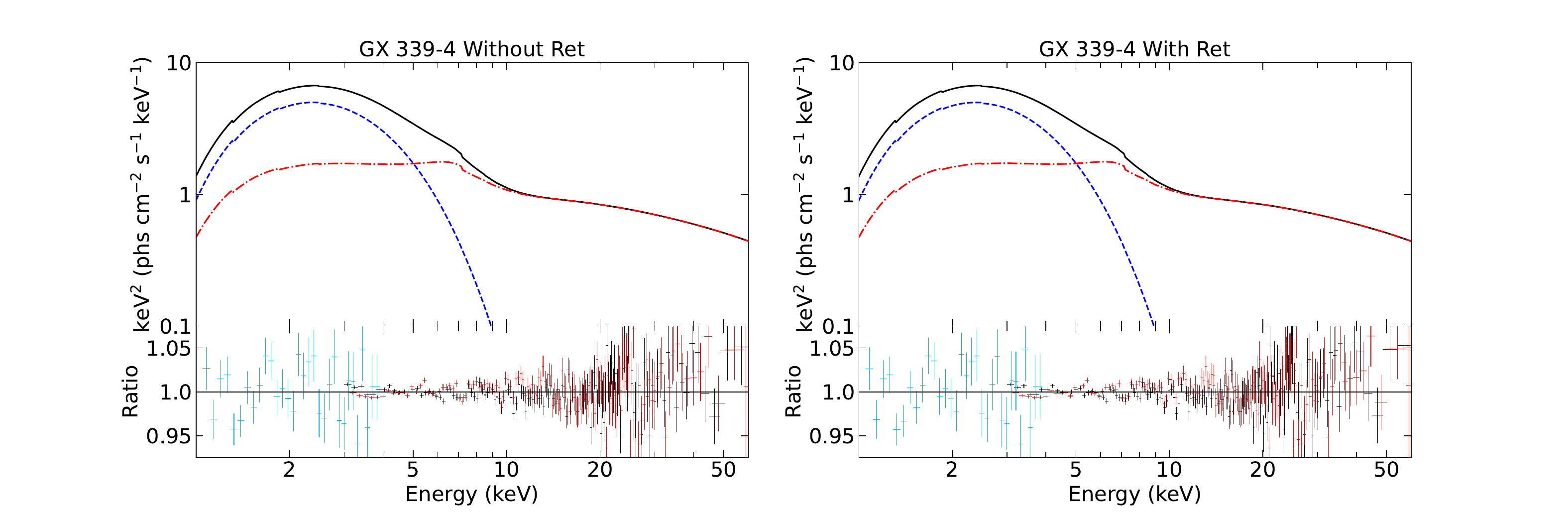}	\vspace{0.5cm}
\end{figure*}

\begin{figure*}
	\centering
	\caption{Best-fit model and residuals for Swift~J1658.2--4242 when we turn the returning radiation off (left panel) and on (right panel). The blue lines represent the total model, the black dashed lines are for the reflection spectrum from the disk and the continuum from the corona ({\tt relxilllpCp}), and the red dashed-dotted lines are to describe a narrow emission line ({\tt gaussian}). In the lower quadrants, the black and red crosses are for \textit{NuSTAR}/FPMA and \textit{NuSTAR}/FPMB data, respectively. }
	\label{swift_model_ra_plot}
	\vspace{0cm}
	\includegraphics[width=0.99\linewidth]{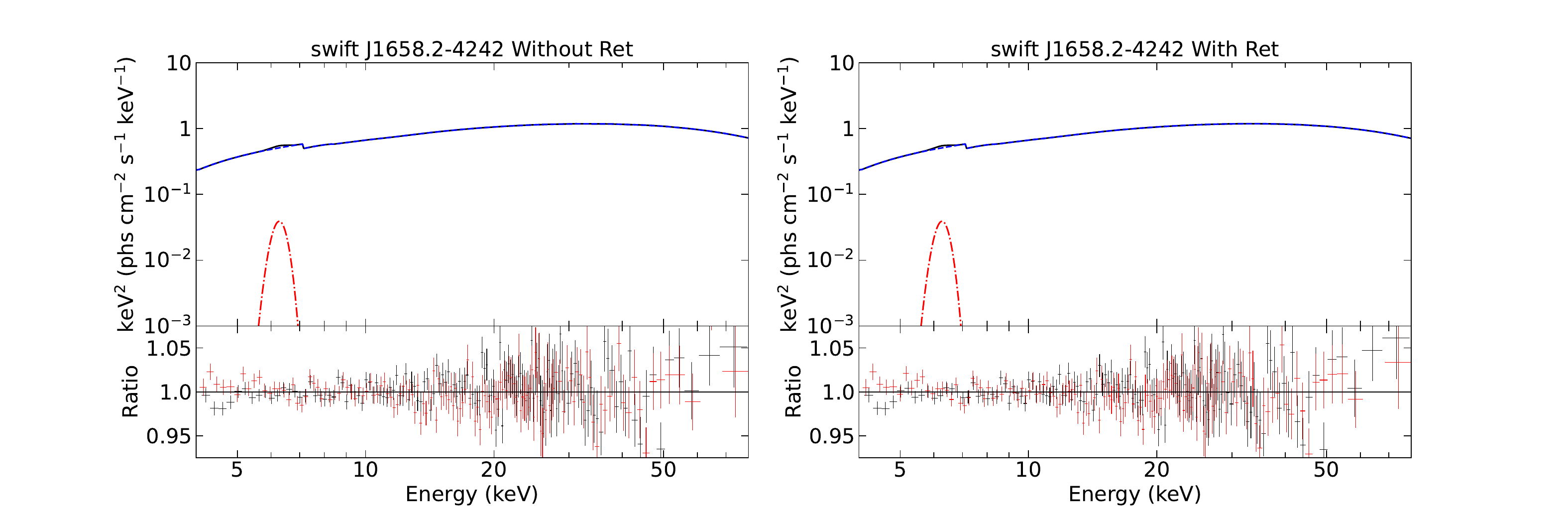}
\end{figure*}

\begin{figure*}
	\centering
	\caption{Best-fit model and residuals for MAXI~J1535--571 when we turn the returning radiation off (left panel) and on (right panel). The black lines represent the total model, the blue dotted lines are for the thermal component ({\tt diskbb}), the red dashed lines are for the reflection spectrum from the disk and the continuum from the corona ({\tt relxilllpCp}), and the orange dashed-dotted lines are the non-relativistic reflection component from distant material ({\tt xillverCp}). In the lower quadrants, the black and red crosses are for \textit{NuSTAR}/FPMA and \textit{NuSTAR}/FPMB data, respectively.}
	\label{MAXI_model_ra_plot}
	\vspace{0cm}
	\includegraphics[width=0.99\linewidth]{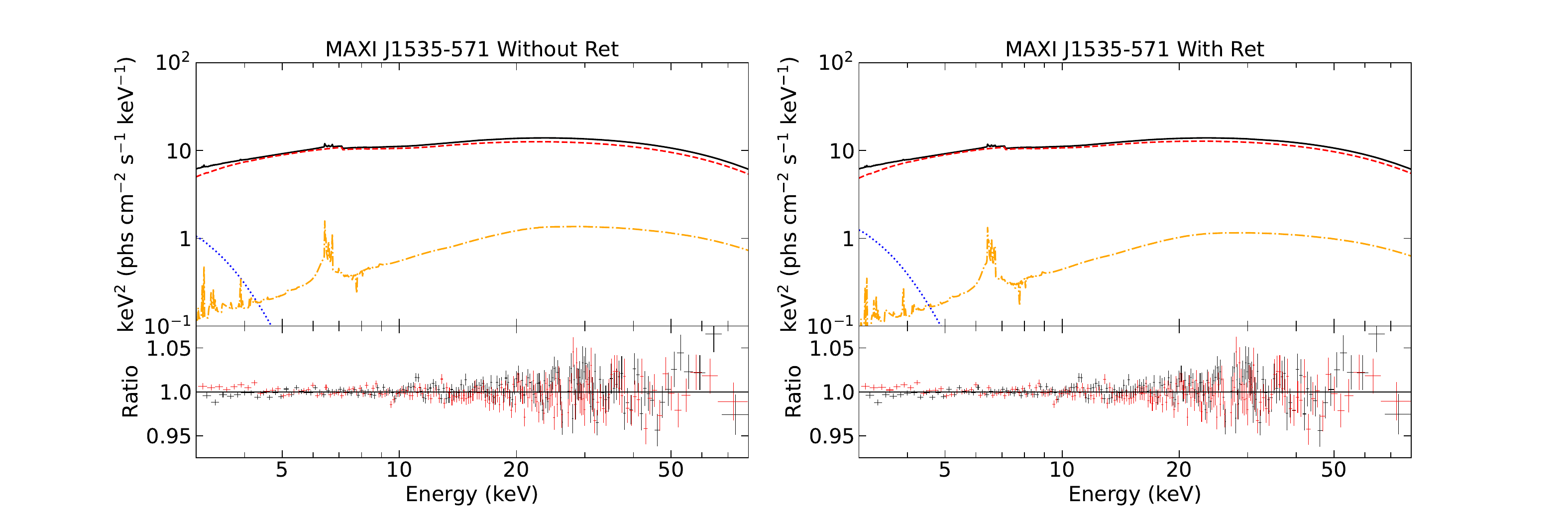}	\vspace{0.5cm}
\end{figure*}

The results of our fits are reported in Tables~\ref{gx339_table}, \ref{swift_table}, and \ref{MAXI_table} for GX~339--4, Swift~J1658.2--4242, and MAXI~J1535--571, respectively. The uncertainties on the best-fit parameters are derived using the {\tt error} command in XSPEC and correspond to 90\% confidence intervals. When we do not report an upper/lower uncertainty, it means that the best-fit value is stuck at one of the boundaries of the parameter range. Figs.~\ref{gx339_model_ra_plot}, \ref{swift_model_ra_plot}, and \ref{MAXI_model_ra_plot} show the best-fit model and the residuals (left panels are without returning radiation and right panels are with returning radiation) for, respectively, GX~339--4, Swift~J1658.2--4242, and MAXI~J1535--571. The discussion of our results is postponed to the next section.

\begin{table*}
\centering
\renewcommand\arraystretch{1.5}
\caption{Best-fit values from the analysis of the spectrum of GX~339--4. The reported uncertainties correspond to the 90\% confidence level for one relevant parameter. When there is no upper/lower uncertainty, the best-fit value is stuck at one of the boundaries of the parameter range.}
\label{gx339_table}
\vspace{0.3cm}
\begin{tabular}{lcc}

\hline

Parameter & Without & With \\
& \hspace{0.3cm} returning radiation \hspace{0.3cm} & \hspace{0.3cm} returning radiation \hspace{0.3cm} \\

\hline

{\tt tbabs} \\
$N_{\rm H}$ [$10^{22}$~cm$^{-2}$] & $0.982_{-0.023}^{+0.023}$ & $0.982_{-0.023}^{+0.023}$\\
\hline

{\tt kerrbb} \\

$M$ [$M_\odot$] & $12.0_{-0.9}^{+2.1}$ & $12.1_{-0.9}^{+1.2}$\\

$\dot M$ [$10^{18}$~g~s$^{-1}$] & $0.78_{-0.17}^{+1.6}$ & $0.80_{-0.16}^{+2.3}$\\

$D$ [kpc] & $9.0_{-2.9}^{+2.4}$ & $9.1_{-2.7}^{+9}$\\
\hline

{\tt relxilllpCp} \\

$\Gamma$ & $2.226_{-0.007}^{+0.009}$ & $2.227_{-0.007}^{+0.009}$\\

$kT_{\rm e}$ [keV] & $400_{-94}$ & $400_{-92}$\\

$h$ [$r_{\rm g}$] & $7.5_{-0.3}^{+0.6}$ & $7.6_{-0.3}^{+0.6}$\\

$a_{*}$ & $0.998_{-0.014}$ & $0.998_{-0.014}$\\

$i$ [deg] & $22.1_{-1.0}^{+0.9}$ & $22.0_{-1.4}^{+1.1}$\\

A$_{\rm Fe}$ & $10.0_{-0.5}$ & $10.0_{-0.5}$\\

$R_{\rm f}$ & $0.91_{-0.17}^{+0.23}$ & $0.77_{-0.13}^{+0.18}$\\

$\log N$ [$N$ in $10^{15}$~cm$^{-3}$] & $17.57_{-0.10}^{+0.11}$ & $17.56_{-0.06}^{+0.11}$\\

$\log\xi$ [$\xi$ in erg~cm~s$^{-1}$] & $4.35_{-0.04}^{+0.04}$ & $4.35_{-0.04}^{+0.04}$\\

Norm & $0.023_{-0.004}^{+0.003}$ & $0.023_{-0.003}^{+0.003}$\\
\hline

$C_{\rm FPMB}$ & $0.9820_{-0.0011}^{+0.0011}$ & $0.9820_{-0.0011}^{+0.0011}$\\

$C_{\rm XRT}$ & $0.95_{-0.01}^{+0.01}$ & $0.95_{-0.01}^{+0.01}$\\

\hline

$\chi^2$/dof & \tabincell{c}{644.54/442=\\1.45824} & \tabincell{c}{645.98/442=\\1.46149}\\


\hline

\end{tabular}

\end{table*}

\begin{table*}
\centering
\renewcommand\arraystretch{1.5}
\caption{Best-fit values from the analysis of the spectrum of Swift~J1658.2--4242. The reported uncertainties correspond to the 90\% confidence level for one relevant parameter. When there is no upper/lower uncertainty, the best-fit value is stuck at one of the boundaries of the parameter range.}
\label{swift_table}
\vspace{0.3cm}
\begin{tabular}{lcc}

\hline

Parameter & Without & With \\
& \hspace{0.3cm} returning radiation \hspace{0.3cm} & \hspace{0.3cm} returning radiation \hspace{0.3cm} \\

\hline

{\tt tbabs} \\
$N_{\rm H}$ [$10^{22}$~cm$^{-2}$] & $18.1_{-0.5}^{+0.4}$ & $17.9_{-0.5}^{+0.5}$\\
\hline
{\tt relxilllpCp} \\
$\Gamma$ & $1.664_{-0.018}^{+0.022}$ & $1.651_{-0.014}^{+0.019}$\\

$kT_{\rm e}$ [keV] & $58.3_{-5}^{+5}$ & $57.1_{-4}^{+6}$\\

$h$ [$r_{\rm g}$] & $2.0^{+0.6}$ & $2.0^{+0.3}$\\

$a_{*}$ & $0.998_{-0.13}$ & $0.998_{-0.11}$\\

$i$ [deg] & $64.4_{-1.7}^{+1.3}$ & $63.0_{-1.9}^{+1.6}$\\

A$_{\rm Fe}$ & $0.50^{+0.08}$ & $0.50^{+0.07}$\\

$R_{\rm f}$ & $1.7_{-0.6}^{+0.7}$ & $1.02_{-0.18}^{+0.22}$\\

$\log N$ [$N$ in $10^{15}$~cm$^{-3}$] & $15.0^{+1.6}$ & $15.0^{+1.5}$\\

$\log\xi$ [$\xi$ in erg~cm~s$^{-1}$] & $3.14_{-0.14}^{+0.13}$ & $3.19_{-0.16}^{+0.08}$\\

Norm & $0.097_{-0.021}^{+0.022}$ & $0.096_{-0.015}^{+0.07}$\\
\hline
{\tt gaussian} \\
$E_{\rm line}$ & $6.25_{-0.04}^{+0.04}$ & $6.24_{-0.04}^{+0.04}$\\

$\sigma$ & $0.25_{-0.06}^{+0.06}$ & $0.25_{-0.06}^{+0.06}$\\

Norm & $7.7_{-1.6}^{+0.9}\times10^{-4}$ & $7.8_{-1.4}^{+1.4}\times10^{-4}$\\
\hline
$C_{\rm FPMB}$ & $0.9867_{-0.0025}^{+0.0025}$ & $0.9867_{-0.0025}^{+0.0025}$\\

\hline

$\chi^2$ /dof & \tabincell{c}{557.92/474=\\1.17705} & \tabincell{c}{559.59/474=\\1.18057}\\


\hline

\end{tabular}

\end{table*}

\begin{table*}
\centering
\renewcommand\arraystretch{1.5}
\caption{Best-fit values from the analysis of the spectrum of MAXI~J1535--571. The reported uncertainties correspond to the 90\% confidence level for one relevant parameter. When there is no upper/lower uncertainty, the best-fit value is stuck at one of the boundaries of the parameter range. $P$ indicates that the 90\% confidence level reaches the boundary of the parameter range.}
\label{MAXI_table}
\vspace{0.3cm}
\begin{tabular}{lcc}

\hline

Parameter & Without & With \\
& \hspace{0.3cm} returning radiation \hspace{0.3cm} & \hspace{0.3cm} returning radiation \hspace{0.3cm} \\

\hline

{\tt tbabs} \\
$N_{\rm H}$ [$10^{22}$~cm$^{-2}$] & $7.0_{-0.6}^{+0.7}$ & $7.1_{-0.8}^{+0.8}$\\
\hline
{\tt diskbb} \\
$T_{\rm in}$ [keV] & $0.410_{-0.03}^{+0.029}$ & $0.424_{-0.04}^{+0.024}$\\

Norm [$10^{4}$] & $8.3_{-2.8}^{+7}$ & $7.4_{-2.2}^{+5}$\\
\hline

{\tt relxilllpCp} \\
$\Gamma$ & $1.794_{-0.011}^{+0.008}$ & $1.790_{-0.006}^{+0.009}$\\

$kT_{\rm e}$ [keV] & $65.8_{-4}^{+14.3}$ & $64.9_{-2.6}^{+7}$\\

$h$ [$r_{\rm g}$] & $2.0^{+0.3}$ & $2.00^{+0.22}$\\

$a_{*}$ & $0.928_{-0.06}^{+0.024}$ & $0.933_{-0.06}^{+0.027}$\\

$i$ [deg] & $55.2_{-1.2}^{+4}$ & $56.6_{-1.1}^{+1.6}$\\

A$_{\rm Fe}$ & $0.73_{-0.05}^{+0.19}$ & $0.84_{-0.12}^{+0.29}$\\

$R_{\rm f}$ & $1.4_{-0.3}^{+0.5}$ & $1.05_{-0.21}^{+0.29}$\\

$\log N$ [$N$ in $10^{15}$~cm$^{-3}$] & $20.0_{-1.2}$ & $19.2_{-0.4}^{+P}$\\

$\log\xi$ [$\xi$ in erg~cm~s$^{-1}$] & $3.23_{-0.06}^{+0.26}$ & $3.41_{-0.3}^{+0.09}$\\

Norm & $1.7_{-0.9}^{+0.4}$ & $1.6_{-0.4}^{+0.7}$\\
\hline

{\tt xillverCp} \\
$\log\xi$ [$\xi$ in erg~cm~s$^{-1}$] & $2.16_{-0.06}^{+0.11}$ & $2.30_{-0.16}^{+0.07}$\\

Norm & $0.019_{-0.004}^{+0.004}$ & $0.016_{-0.003}^{+0.005}$\\
\hline

$C_{\rm FPMB}$ & $0.9481_{-0.0009}^{+0.0009}$ & $0.9481_{-0.0009}^{+0.0009}$\\

\hline

$\chi^2$ /dof & \tabincell{c}{659.73/520=\\1.26871} & \tabincell{c}{661.22/520=\\1.27158}\\


\hline

\end{tabular}

\end{table*}

\section{Discussion and conclusions}\label{discussion}

In this study, we have explored the impact of the returning radiation on the spectral fitting of X-ray data from three Galactic black holes: GX~339--4, Swift~J1658.2--4242, and MAXI~J1535--571. Our primary objective was to determine whether returning radiation significantly influences the estimation of key model parameters, such as black hole spin, primary source height, and iron abundances. Based on our analysis, we observed minimal differences between the results obtained from models including and excluding returning radiation.

For GX~339--4, the system has a very high black hole spin parameter and a moderately high value of the height of the corona ($h \sim 7.5$~$r_{\rm g}$), but the quality of the data is very good and therefore the measurement of most parameters of the system is quite precise. We do not find any clear difference in the estimate of the parameters of the source if we include or we do not include the effect of the returning radiation in {\tt relxilllpCp}. In the case of Swift~J1658.2--4242, we find both an extremely high value of the black hole spin parameter ($a_* > 0.98$) and a very low value of the height of the corona ($h < 3$~$r_{\rm g}$), making this source an ideal candidate to test the impact of the returning radiation on the estimate of the model parameters. Despite the promising properties of this source, we find that the impact of the returning radiation is still marginal and the estimates of the black hole spin parameter, coronal height, iron abundance, and the other parameters are consistent between the models without and with returning radiation. Last, we have analyzed a spectrum of MAXI~J1535--571: for this source, the black hole spin parameter is still high, even if not extremely high as in the case of GX~339--4 and Swift~J1658.2--4242, and the estimate of the height of the corona is very low ($h < 3$~$r_{\rm g}$). Once again, we find consistent measurements of the model parameters between the fits without and with returning radiation in {\tt relxilllpCp}.

According to \citet{2022MNRAS.514.3965D}, if we ignore the returning radiation we may overestimate the coronal height, $h$, and the reflection fraction, $R_{\rm f}$. In our analyses, we do not see any clear difference in the estimate of the height of the corona. Concerning the estimate of the reflection fraction, we confirm that the model with returning radiation provides a lower best-fit value of the reflection fraction, but the measurements without and with returning radiation are still marginally consistent if we consider their uncertainties at the 90\% confidence level.

Last, we note that in {\tt relxill} the effect of the returning radiation is considered only in the emissivity profile of the lamppost corona (i.e., how the corona illuminates the disk) and the non-relativistic reflection spectrum is still calculated assuming that the incident radiation is a Comptonized spectrum (while the spectrum of the incident radiation should be the combination of the Comptonized spectrum from the corona and the spectrum of the returning radiation). As shown in Ref.~\cite{2024ApJ...965...66M}, such an approximation is not supposed to work well for very high values of the black hole spin parameters and low values of the coronal height, so in the case of our sources (at least for Swift~J1658.2--4242 and MAXI~J1535--571, where $h < 3$~$r_{\rm g}$). At present, we cannot fit the \textit{NuSTAR} spectra of our sources with the model presented in Refs.~\cite{2024ApJ...965...66M,2024ApJ...976..229M} because that model is too slow to analyze observations. It is still possible that with that model we can obtain different estimates of some parameters. This will be checked after developing a faster version of the model in Refs.~\cite{2024ApJ...965...66M,2024ApJ...976..229M} suitable for the analysis of real data.

\section*{Data Availability Statement}

The data analyzed in this article are openly available~\cite{HEASARC}.


\begin{acknowledgments}
This work was supported by the Natural Science Foundation of Shanghai, Grant No.~22ZR1403400, and the National Natural Science Foundation of China (NSFC), Grant No.~12250610185 and Grant No.~12261131497.
\end{acknowledgments}

\bibliography{bibliography}

\end{document}